# Microstructural Effects of Chemical Island Templating in Patterned Matrix-Pillar Oxide Nanocomposites

R.B. Comes,[a+] K. Siebein,[b] J. Lu,[a] and S.A. Wolf [a, c]

The ability to pattern the location of pillars in epitaxial matrix-pillar nanocomposites is a key challenge to develop future technologies using these intriguing materials. One such model system employs a ferrimagnetic $CoFe_2O_4$ (CFO) pillar embedded in a ferroelectric $BiFeO_3$ (BFO) matrix, which has been proposed as a possible memory or logic system. These composites self-assemble spontaneously with pillars forming through nucleation at a random location when grown via physical vapor deposition. Recent results have shown that if an island of the pillar material is pre-patterned on the substrate, it is possible to control the nucleation process and determine the locations where pillars form. In this work, we employ electron microscopy and x-ray diffraction to examine the chemical composition and microstructure of patterned CFO-BFO nanocomposites. Cross-sectional transmission electron microscopy is used to examine the nucleation effects at the interface between the template island and resulting pillar. Evidence of grain boundaries and lattice tilting in the templated pillars is also presented and attributed to the microstructure of the seed island.

## I. Introduction

Epitaxial matrix-pillar oxide nanocomposites offer unique opportunities for future applications in spintronic logic[1] and memory[2] devices. By selecting suitable complex oxides, a wide variety of functionalities can be achieved, such as multiferroic properties with a ferroelectric matrix and ferromagnetic pillar[3–6], magneto-optical properties through a photostrictive matrix and magnetoelastic pillar[7], or a metallic Fe nanowire embedded in anti-ferromagnetic $LaSrFeO_4$ matrix, which has potential for magnetic exchange bias.[8] Novel functionalities in a wide variety of epitaxial nanocomposites is often observed due to the misfit strain between the matrix and pillar.[9] These epitaxial composites, also referred to as vertically-aligned nanocomposites, typically form through spontaneous self-assembly during film growth, which is the result of immiscibility between the two material systems. Such results are particularly common in spinel-perovskite nanocomposites.[10] These two oxide crystal systems have different phase and surface energies, making it energetically favorable to phase segregate and produce a composite system. When nanocomposite films are grown epitaxially on (001)-oriented perovskite substrates such as $SrTiO_3$ (STO), the minimum energy configuration occurs when an epitaxial spinel pillar forms in an epitaxial perovskite matrix, with cube-on-cube epitaxy with the substrate for both the pillar and matrix.

Nanocomposites comprised of $CoFe_2O_4$ (CFO) pillars and a $BiFeO_3$ (BFO) matrix have attracted particular interest because CFO is a ferrimagnetic spinel which exhibits strong magnetoelastic response,[11] while BFO is a ferroelectric with a large piezoelectric $d_{33}$ coefficient[12]. Most results in the literature have demonstrated that due to the lattice mismatch between CFO and BFO, which is about 4 % along the out-of-plane direction, residual strains in the pillar are present after growth and produce perpendicular magnetic anisotropy.[2,13] However, the origin of this strain is open for debate, as one group has shown in the similar $NiFe_2O_4$-BFO composite system that residual strain is entirely relaxed.[14] Others have suggested that residual strain in the system may be attributed to differences in thermal expansion in CFO, BFO and the commonly used STO substrates.[15] For CFO-BFO composites with perpendicular magnetic anisotropy, it has been shown that the application of an electric field to the BFO matrix will induce a strain in the CFO pillar and reduce the magnetic anisotropy.[2,16] These exciting results have led to the proposal of a reconfigurable magnetic logic architecture made up of a CFO-BFO nanocomposite that has been patterned to produce a square array of CFO pillars with periodicity of 100 nm or less.[1,17] However, the ability to pattern epitaxial nanocomposites has thus far proven challenging, which is the focus of this work. Previous results have shown that it is possible to direct the self-assembly of CFO-BFO nanocomposite by patterning CFO islands on the surface of a Nb-doped STO (Nb:STO) substrate.[18]

An alternative method using a liftoff technique to pattern CFO islands was recently reported, showing similar results for CFO-BFO nanocomposites.[19] Patterning of pits on the substrate using a focused ion beam (FIB) system may also be used to induce nucleation of the CFO islands, followed by nanocomposite growth.[20] These works demonstrated that CFO pillars will form at the template island sites, with the island sites collecting all CFO flux if it is kinetically possible to do so. The multiferroic properties of these composites were also examined and shown to be comparable to unpatterned composites. Subsequent work by others on Fe-LaSrFeO$_4$ patterned pillar-matrix composites has confirmed the kinetic surface diffusion model.[21,22]

In this work, we examine the structural properties of island-templated CFO-BFO nanocomposites in detail using x-ray diffraction and transmission electron microscopy. An array of CFO islands is fabricated via electron-beam lithography and reactive ion etching and is used to template the growth of the CFO-BFO nanocomposite. The effect of the microstructure of the CFO seed island on the resulting pillar is examined and the results are compared to unpatterned nanocomposites. These results elucidate the structural mechanisms that promote the templated growth of CFO-BFO nanocomposites.

## II. Experimental Methods

For all oxide films, pulsed electron deposition (PED) was employed.[23,24] The system is equipped with two electron guns, which are used to separately ablate a Bi$_{1.15}$FeO$_3$ and CoFe$_2$O$_4$ target for film growth. To produce a template substrate, an initial uniform CFO film was grown epitaxially on Nb:STO. The growth nanocomposite films have been described in previously.[25] The significant lattice mismatch and differences in surface energy lead to the formation of CFO islands on the surface via the Volmer-Weber epitaxial growth mode.[26] The island grain size is uniform across the surface, with diameters between 25 and 50 nm, and the film thickness was measured to be approximately 12.5 nm via x-ray reflectivity. The CFO film was then patterned via electron-beam lithography and etched using reactive ion etching to produce square arrays of CFO template islands with nominal periodicity of 100 nm between islands. Details of this fabrication process can be found elsewhere.[18] An atomic force microscopy image of the resulting array is shown in Figure 1. The peak of the island is generally between 4 nm and 5 nm above the exposed substrate between the islands. Away from the patterned array, the Nb:STO substrate shows an extremely smooth surface with step-edges visible from the initial chemical surface treatment and annealing process, indicating a high surface quality for subsequent film growth.

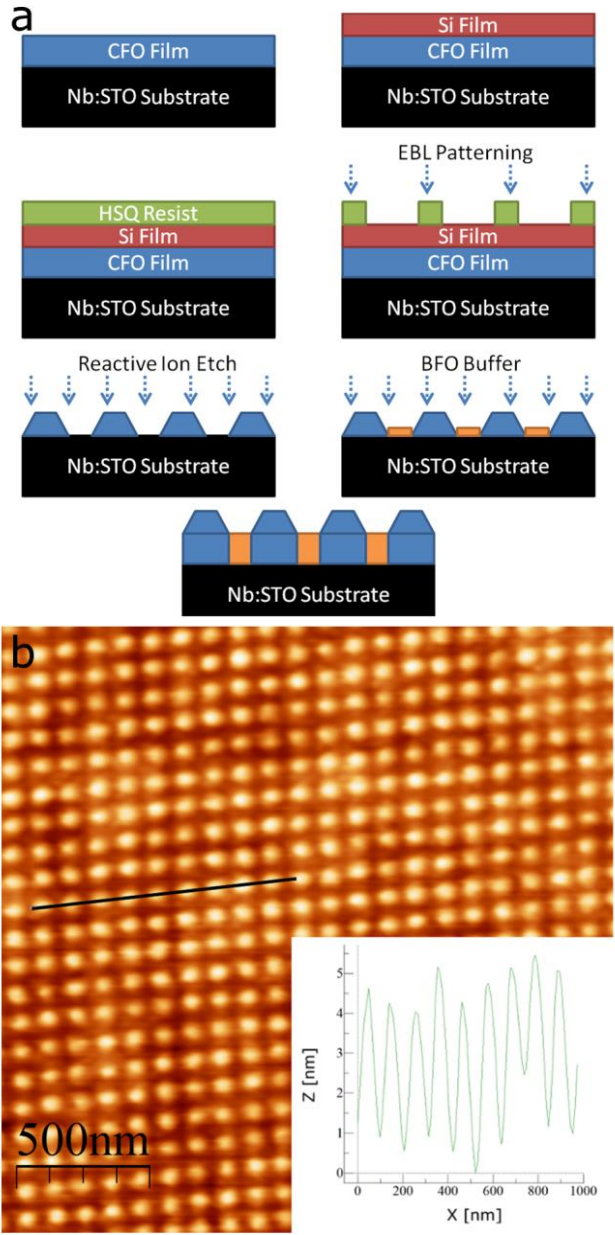

**Figure 1.** a) Fabrication process schematic; b) Atomic force microscopy topography map of template island array patterned for composite growth; Inset) Height profile along black line in image. Black line is parallel to [100] direction.

The patterned sample was then loaded back into the PED system, where a second deposition was performed to grow a CFO-BFO nanocomposite. An initial BFO layer calibrated to be 1 nm thick was grown to wet the surface of the substrate and prevent nucleation of CFO pillars away from the template sites. The lattice and crystal symmetry mismatch between the perovskite BFO and spinel CFO leads to preferential diffusion of BFO flux off of the CFO island sites,[27] leaving them exposed for subsequent growth. The second electron gun was then activated and a composite film was grown by co-depositing from the CFO and BFO targets, with the total BFO matrix thickness calibrated to be approximately 25 nm. Subsequent x-ray reflectivity measurements confirmed that the total BFO matrix thickness was 24.9 nm. The area density of CFO pillars was calibrated to be approximately 10 % of the overall surface.

To characterize the templating effects of the CFO island on the resulting nanocomposite, the sample was characterized by x-ray diffraction (XRD), scanning electron microscopy (SEM), transmission electron microscopy (TEM) and high-angle annular dark field scanning transmission electron microscopy (HAADF-STEM) with energy dispersive x-ray spectroscopy (EDS). Details of these data analysis techniques can be found in the supplementary online information. A single array with pitch of 100 nm along the [100] in-plane directions was used for all measurements. These measurements were useful to determine the epitaxial configuration of the pillars, which are expected to have {110}-type in-plane facets with the BFO matrix.[4] The cross-sectional TEM sample was prepared via a standard focused ion beam liftout and thinning process. The sample was cross-sectioned along the <110> in-plane direction to align with the faceting structure of the CFO pillar. A total of ~10 pillars in the lamella over an ~1.5 μm length were thin enough to examine in detail with the TEM.

## III. Experimental Results and Discussion

### 1. BiFeO$_3$ Matrix

The sample was characterized via out-of-plane XRD and a reciprocal space map (RSM) about the Nb:STO (103) peak. These results are shown in Figure 2(a-b). Due to the small area coverage of the template across the entire area of the substrate (10 μm x 10 μm array on a 5 mm x 5 mm substrate), the measurements are based almost entirely on the structural properties of the film away from the patterned region. Electron microscopy is required to examine the structural properties of the patterned arrays. XRD measurements were used to determine the out-of-plane pseudocubic (pc) lattice parameter of the BFO film, which was found to be 4.07±0.01 Å using Cu Kα$_1$ radiation with a wavelength of 1.5046 Å with a (002) peak at 44.45±0.05°. The lattice parameter of any CFO pillars could not be determined due to the low area density of the CFO phase. No statistically significant CFO peak was observed above the noise level of the data. The close overlap between the location of the bulk CFO (004) diffraction peak (highlighted in Figure 2(a)) and the BFO (002)$_{pc}$ diffraction peak masks any signal that might be present. The RSM, which is shown in Figure 2(b), indicated that the BFO film was coherent with the Nb:STO substrate along the in-plane direction. The reciprocal lattice coordinate, $Q_x$, represents the diffraction peak along the [100] in-plane direction, while the $Q_z$ coordinate represents the out-of-plane [001] direction. Both the BFO and Nb:STO peaks fall at the same value of $Q_x$, indicating coherent strain at the interface. The absence of asymmetric smearing of the BFO peak across towards lower magnitude $Q_x$ values further supports this conclusion, indicating a homogeneous in-plane lattice parameter.[28] This observation across the entire film was confirmed to be valid near the patterned pillars through TEM measurements, which are shown in Figure 3(c-d). The image was Fourier filtered to show the (110) planes of the film and substrate, which are coherent across the interface. Details of this process can be found in the supporting online information.

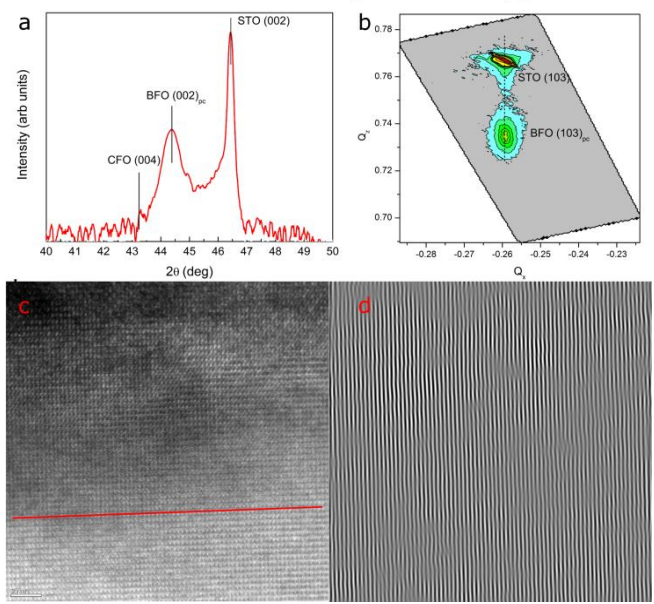

**Figure 2.** Out-of-plane x-ray diffraction (a) and reciprocal space map (b) of sample showing the (103) diffraction peak of the Nb-doped SrTiO$_3$ (Nb:STO) substrate and pseudocubic (103) peak of the BiFeO$_3$ (BFO) matrix. The nominal bulk CoFe$_2$O$_4$ peak angle is noted though the intensity does not rise above background. High-resolution transmission electron micrograph (c) of interface between BFO matrix and Nb:STO substrate with Fourier filtered image (d) showing in plane coherency of matrix and substrate.

### 2. Chemical Templating

Wide view and high resolution SEM images of the patterned array are shown in Figure 3. The horizontal axis of the figure is aligned to correspond to the [100] substrate axis. The pillars are apparently facetted along the {110} planes, indicating an epitaxial interface with the BFO matrix. There is some variation in pillar size from site to site, but there is 100 % pattern fidelity within the image. Over larger regions of the array, which is a 100x100 array covering a 10 μm by 10 μm area, there are some defective regions that are most likely the result of the deposition of particulates during the PED growth process.

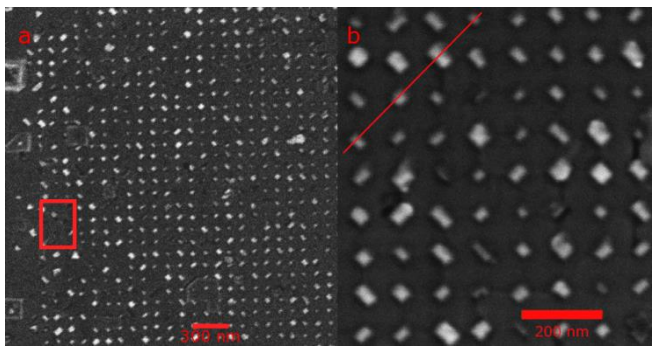

**Figure 3.** Scanning electron microscope images of templated array of CoFe$_2$O$_4$ nanopillars in BiFeO$_3$ matrix with 100 nm pitch. Horizontal axis is parallel to [100] direction. Box shows edge region with missing pillars. a) Wide view with edge of array visible. Scale bar represents 300 nm. b) High resolution view of array. Scale bar represents 200 nm. Diagonal line represents cross section axis along [110] direction for transmission electron microscopy analysis.

Examining the pattern at the edge of the array provides useful information regarding the chemical templating effects of the seed islands. A box highlights one such region in Figure 3(a). During the EBL patterning process, the effective electron dose in the HSQ resist near the edge of the array is reduced due to fewer backscattered electrons from exposure of the nearby islands. When the resist is developed during the fabrication process, the diameter of the pillars near the edge is reduced and the resulting seed islands are smaller. While there is still topographic evidence of the islands in AFM scans, it appears that the CFO seed layer is completely removed in some cases given that pillars are more frequently missing near the edges of the pattern. Further evidence of this phenomenon can be found in Figure S2 of the supplementary online information, which shows an array that received too large of an effective dose in the center of the array to fully remove the CFO layer between template sites. At the corner of the array, however, templated pillars do form due to reduced EBL dose. Collectively, these results show that a CFO chemical seed island, rather than simply a topographic feature on the substrate, is required to nucleate a pillar. It is unknown if there is a critical size of the CFO island to promote pillar formation, as we do not have a sufficiently large sample set of pillars to characterize the effect of seed island size on the resulting formation. Furthermore, the STO substrate must be exposed away from the pillars to produce an ideal BFO matrix. Figure 3(b) shows the cross-sectioning axis used for liftout for TEM analysis. HAADF-STEM images of two template pillars that were characterized in detail are shown in Figure 4. The spacing between the center of the pillars is approximately 140 nm, which is approximately equal to $\sqrt{2} \times 100$ nm, the spacing that results from extracting the sample along the diagonal. The sensitivity of HAADF measurements to the atomic number, Z, means that Bi, which has atomic number 83 will produce the brightest contrast, while Sr (Z = 38), Ti (Z = 22), Co (Z = 27), and Fe (Z = 26), will be darker. There is a brighter contrast around the portion of the pillar that is below the surface of the matrix, which can be attributed to residual BFO in front of or behind the pillar. Bright spots on the film and pillar surface can be attributed to re-deposition of high Z elements, such as Bi or the Pt protective layer, during the FIB milling process. The thickness of the cross-section sample is estimated to be between 50 nm and 75 nm—greater than the width of the pillar shown in the SEM image in Figure 3.

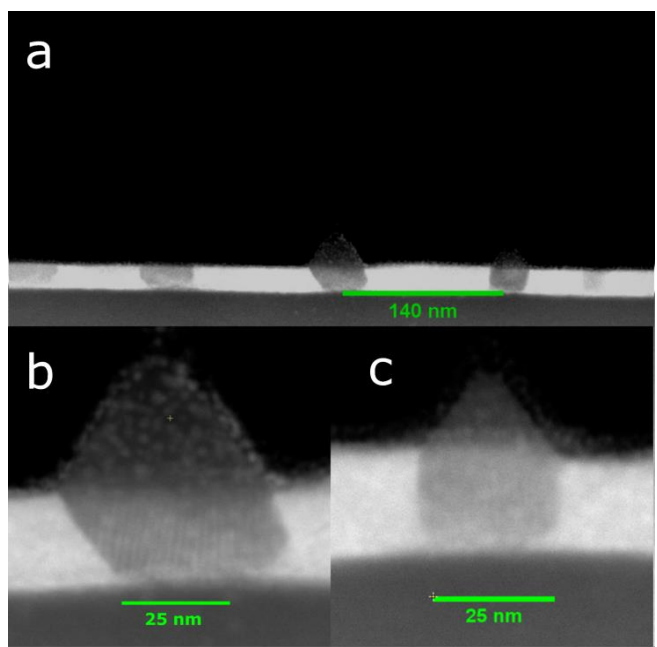

**Figure 4.** HAADF-STEM image of patterned nanocomposite. a) Wide image showing uniform spacing of pillars; b) Image of left pillar in (a); c) Image of right pillar in (a).

High resolution HAADF images of the two pillars of interest are shown in Figure 4(b-c). The base of the pillar is approximately 25 nm in width in both cases and {111}-facets are visible above the surface of the matrix, which was confirmed by measurements of the angle between the facet and the horizontal matrix surface. Interestingly, the interfaces beneath the surface for the two pillars are very different. In Figure 4(b), the pillar tilts to one side, while in Figure 4(c) the pillar is vertical. This can most likely be attributed to the difference in nucleation behavior at the surface of the CFO island that was patterned on the substrate. To further understand the nature of the interface, energy dispersive x-ray spectroscopy (EDS) maps were taken for both pillars. In these measurements, a spectrum of x-ray energies is obtained across a uniform grid of points using the STEM mode of the microscope.

The results of the EDS map for the pillar shown in Figure 4(c) are shown in Figure 5, with the acquired HAADF STEM image (a), green map (b) corresponding to the Bi Lα peak, the red map (c) corresponding to the Co Kα peak, and the blue map (d) corresponding to the Ti Kα peak. The peak was acquired over an area of 60 nm by 40 nm, with 2 nm pixel size along both directions. It should be noted that the Co Kα peak is very close in the energy spectrum to the Fe Kβ peak, meaning that the peak at the Co Kα location attributed to the BFO matrix will be non-zero as a result of the Fe present in the matrix. Additionally, the presence of the BFO in front of or behind the pillar means that the Bi map will also have significant intensity in the pillar region. To clarify the results, image processing was performed so that lower intensity regions were partially transparent and a single image with all color maps overlaid was created. The result is shown in (e). An additional EDS map was also collected with 1 nm pixel size at the interface between the pillar and substrate to examine the nucleation effects. The same overlay procedure was performed and the results of this map are shown in (f).

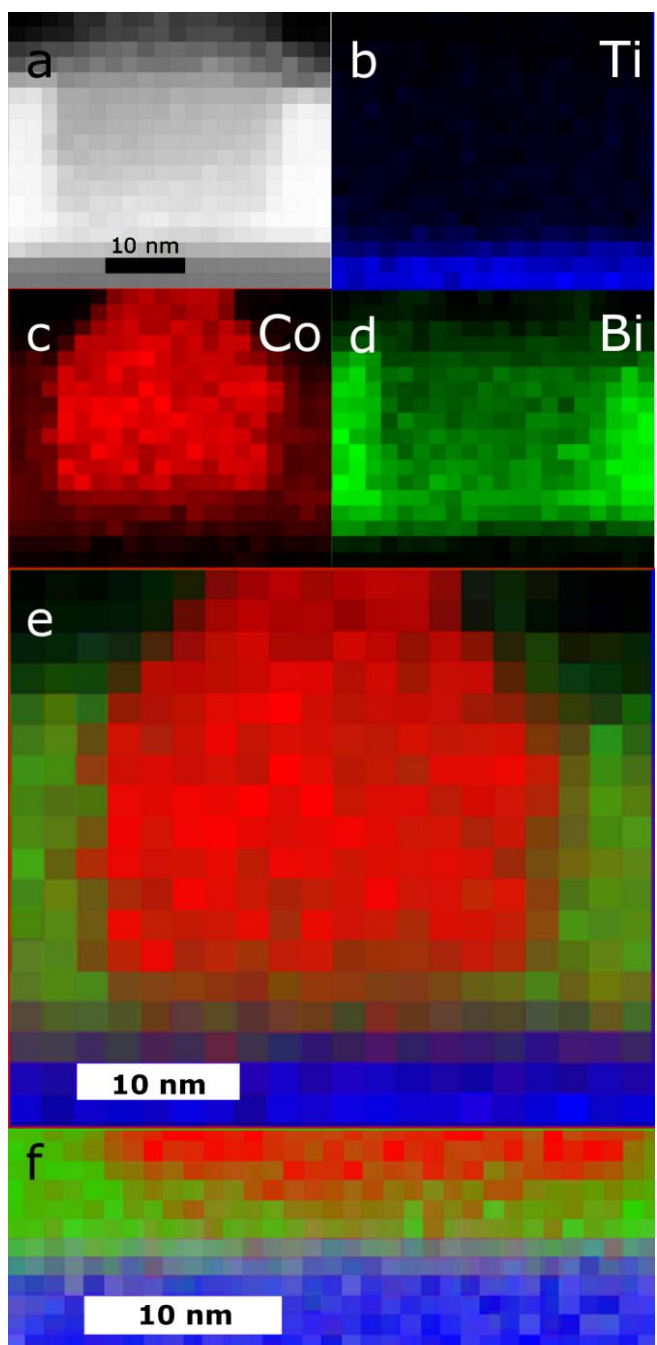

Figure 5. Energy dispersive x-ray spectroscopy map of pillar from Figure 4(c). a) HAADF signal at each pixel measured; b) Ti Kα edge map; c) Co Kα edge map; d) Bi Lα edge map; e) Overlay of EDS map images; f) Overlaid images from EDS map of pillar-substrate interface.

Analysis of the maps in Figure 5 is enlightening to explain the growth process, and supports the hypothesis for the chemical nature of the pillar templating. The Ti map was relatively smooth, with no apparent mound of $SrTiO_3$ present beneath the pillar. This indicates that the reactive ion etching process likely removed at most 1 nm (approximately 3 unit cells) of the substrate based on the pixel size chosen for the scan, leaving at least 3 nm of CFO on the surface of the island to serve as a chemical template. Further analysis of the interface between the substrate and template island will be presented below with TEM images of the lattice. The Bi map shows that there is a greater Bi density within 5 nm of the substrate than in the areas farther away. This can likely be attributed to two features of the growth: the initial 1 nm layer of BFO deposited to coat the surface of the substrate, and the outgrowth of the pillar as it increases in height. The initial BFO layer may partially overcoat the edges of the CFO island, producing a greater density of BFO around the edge. Additionally, the narrow base of the CFO pillar means that there would be additional BFO along the beam path in this region. Both contributions would be expected to increase the Bi intensity at the interface, as we observe.

To further explore the chemical templating effects of the CFO island, a second pillar exhibiting significantly different microstructure and shape was also examined. A series of TEM and EDS images of the pillar at various magnifications is shown in Figure 6. An arrow in Figure 6(a) denotes a region of low intensity in the HRTEM image of the pillar. The pillar is otherwise ideal, with Moire fringing due to the presence of lattice-mismatched BFO and CFO along the beam path. The diameter is significantly greater than that of the previous pillar, with a maximum width of approximately 45 nm compared to 25 nm previously. A higher resolution image shown in Figure 6(b) suggests that the dark region is the Nb:STO substrate protruding into the CFO pillar several nm. This protrusion could be the result of over-etching during the template preparation process, possibly leaving a small amount of CFO on top of a Nb:STO island, rather than the idealized model of a CFO island on a flat Nb:STO substrate. To study this, an EDS map was performed at the interface as in the previous measurements, with an area of 40 nm by 20 nm and pixel size of 2 nm by 1 nm. The results of this EDS map are shown in Figure 6(c-g).

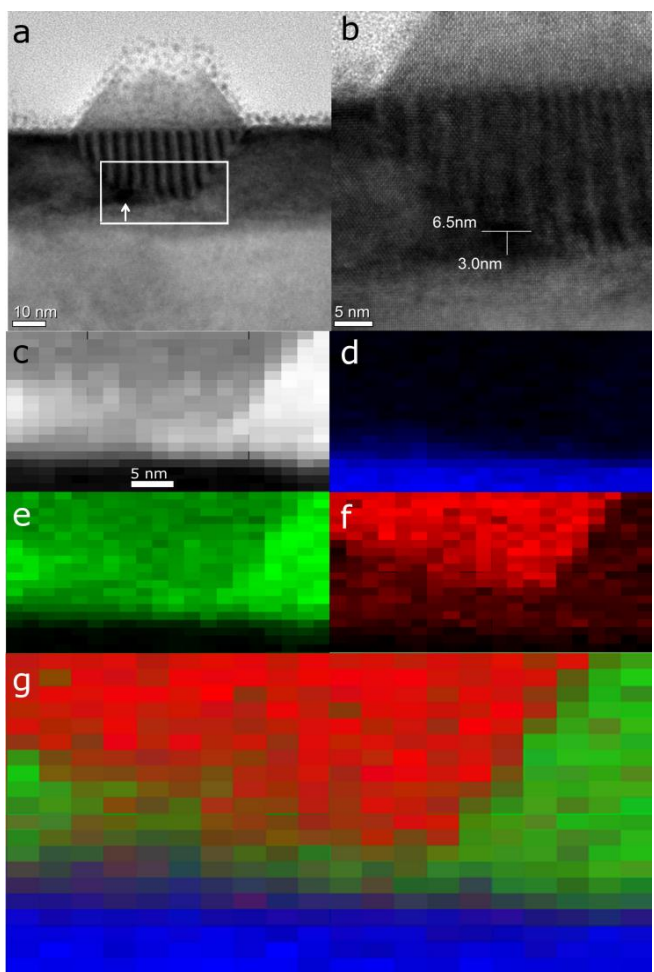

Figure 6. Chemical analysis of pillar. (a) Wide view TEM image showing entire pillar with region of interest for EDS measurements; (b) Narrow view showing region referenced with arrow in (A); (c) STEM image of region examined with EDS; (d) Ti $K_\alpha$ map; (e) Bi $L_\alpha$ map; (f) Co $K_\alpha$ map; (g) Overlaid EDS color maps showing chemical composition at the interface.

The EDS results presented in Figure 6 support the hypothesis based on the TEM images. The Ti $K_\alpha$ map shows a rougher interface, with additional Ti in the region where there is reduced Co intensity. The Bi intensity is in agreement with what would be expected, with higher intensities in areas where there is no Co present and non-zero intensity at the interface due to the thickness of the sample. Measurements of the height of the protruding Nb:STO indicate that it is approximately 3 nm higher than the original surface of the substrate. This is slightly less than the average island height measured in the AFM profile shown in Figure 1, where islands were 4 nm to 5 nm in height. This confirms a small amount of residual CFO was present at the peak of the island to act as a chemical template.

## 3. Microstructural Effects

To explore the effects of the template island on the microstructure of the CFO pillar, several pillars were examined via high-resolution TEM (HRTEM). The pillar from Figure 4(c) and Figure 5 was examined repeatedly after two FIB thinning steps to explore intriguing microstructural features seen in the HRTEM images. Figure 7 shows the pillar after the first FIB thinning step (a) and second (b), with the regions of interest highlighted by white rectangles. The inset of Figure 7(b) shows a HAADF-STEM image of the same pillar with strong contrast between the pillar and matrix, suggesting that there is little or no residual BFO along the beam path after the second FIB thinning step. Irregularities in the HRTEM diffraction contrast can be seen in the highlighted regions of both images, suggesting a complex faulted structure within the pillar. After ~20 nm outward to the vertical surface of the pillar, the irregular features have vanished and uniform contrast is seen near the surface of the BFO matrix and in the truncated pyramid above the surface of the matrix.

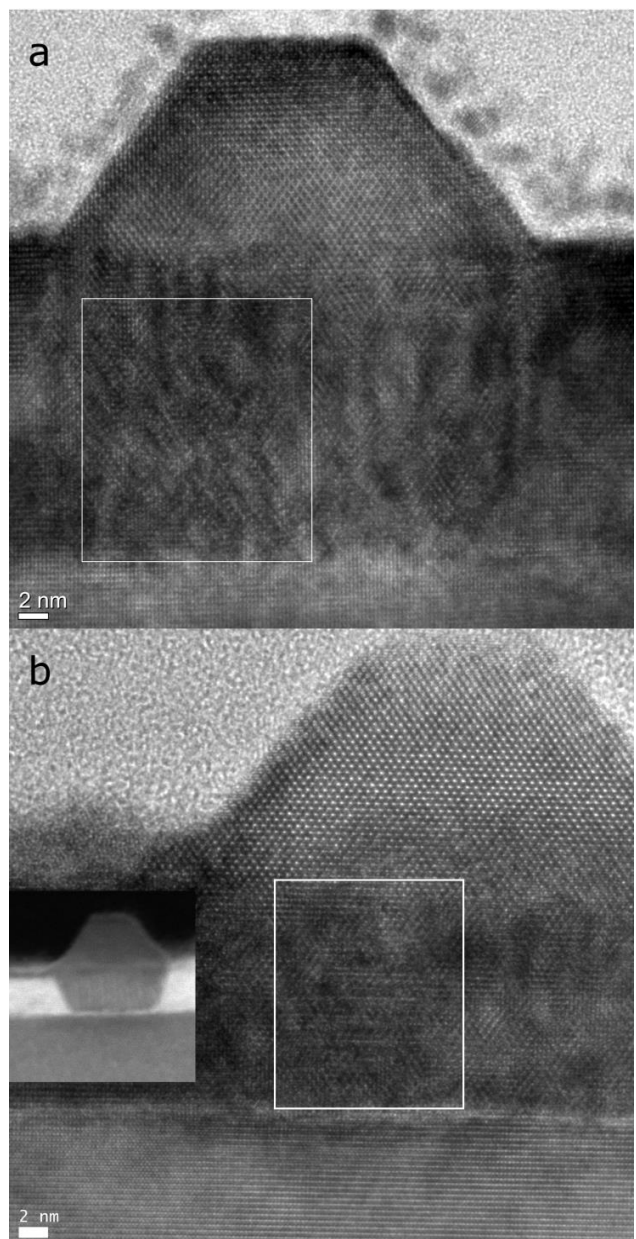

Figure 7. Series of images of same pillar after first (a) and second (b) FIB thinning steps, showing evidence of fault structures within the pillar. White boxes are guides to the eye for regions of interest. Inset of (b) shows HAADF-STEM image of same pillar.

While it is not possible to determine the nature of the fault structure in this pillar given the complex microstructure that seems to be present, the fact that irregular contrast is observed after repeated thinning steps suggests that the contrast is due to features in the CFO pillar rather than any contribution from the BFO matrix. The microstructure can most likely be attributed to the initial structure of the CFO template island. The original epitaxial CFO film on the Nb:STO substrate was grown with the Volmer-Weber island growth mode[25] with coalesced islands producing an epitaxial granular film. This film had island grain diameters between 25 nm and 50 nm. Thus, when e-beam lithography is performed to produce a pillar diameter of 50 nm in the resist, it is very likely that the resulting pillar will cross the grain boundary between two or more islands in the CFO film. Pillars that form from a multi-grained template island would be expected to exhibit a faulted structure similar to what was seen in Figure 7. The frequency at which these multi-grain islands occur would be governed by the ability to fabricate a single grain seed island. Only one pillar in the TEM sample (out of ~10) showed the faulted structure, so a larger sample would be needed to estimate the frequency that these kinds of defects occur. It has been shown in a variety of works that the grain boundary between two epitaxial islands will exhibit intrinsic stresses and dislocations after growth.[29,30] Thus, the resulting pillar will be composed of multiple epitaxial grains during the initial stages of growth. As a means of removing one grain, the dislocations observed could accommodate the expansion of one grain during the growth process at the expense of the other.

The defects observed in this pillar could have a significant effect on the magnetic properties of the pillar. The internal dislocations would be likely to change the strain the pillar, affecting the magnetic anisotropy of the pillar. Additionally, the presence of dislocations or antiphase boundaries in CFO has been shown to reduce the saturation magnetization of the material and reduce the energy barrier to form magnetic domain walls.[31–33] Thus, the structural defects that are present in this pillar could negatively affect the performance of the CFO-BFO nanocomposite in a memory or logic device. Further studies are needed to determine the best means to produce uniform microstructure in pillars over a large pattern area.

Examinations of a second pillar via HRTEM after the second thinning step provided additional insights into the growth process. Figure 8 shows a templated pillar (a) along with a selected area Fast Fourier transform (FFT) of a square region across the CFO-Nb:STO interface (b). Clear diffraction peaks for both the pillar and substrate are visible, with the CFO peaks closer to the central (000) spot due to the larger lattice parameter of CFO compared to STO (8.38 Å/2 = 4.19 Å vs. 3.905 Å). Lines connecting two sets of symmetric diffraction peaks for both CFO and STO are shown in red (STO) and green (CFO). The horizontal lines connect the STO <220> peaks and the CFO <440> peaks, while the diagonal lines connect the STO <444> and CFO <888> peaks. It should be noted that the lines for the STO <220> and CFO <440> peaks are not strictly parallel, as would be expected for cube-on-cube epitaxy. This indicates that there is some lattice tilting at the interface between the pillar and substrate. To our knowledge, this form of tilting has not been reported previously in either patterned or unpatterned spinel-perovskite nanocomposites. The tilting could be attributed to the complex nature of the pillar nucleation on the seed island. The lines for STO <444> and CFO <888> are also not parallel. Normally such an observation would suggest that uniaxial strain is present in the CFO pillar, but the observed lattice tilting could also explain this result. Further studies using another technique such as synchrotron nanodiffraction on a large array of pillars could further explain this observation.

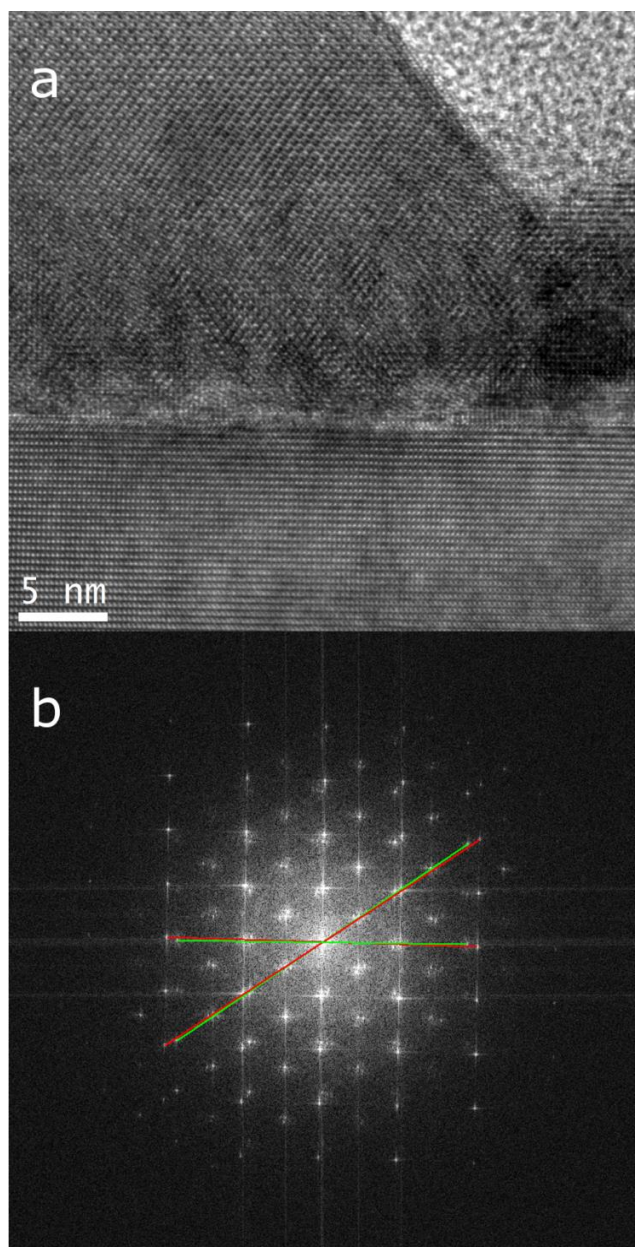

Figure 8. HRTEM image (a) and selected-area FFT (b) of templated pillar with non-parallel red (STO) and green (CFO) lines indicating the tilting of the pillar lattice.

## Conclusions

In summary, we have examined the microstructure of CFO pillars formed in a CFO-BFO nanocomposite using a CFO island template. X-ray diffraction and transmission electron microscopy (TEM) measurements have confirmed that the BFO matrix is coherent to the substrate, in spite of the substrate surface damage that might have been expected from the Ar ion etch to remove the initial CFO film. We have also shown that the patterned CFO island promotes the growth of the pillar through a chemical nucleation process and that topographic changes alone are not sufficient to promote nucleation. The fabrication technique to produce the island template has also been shown to have an effect on the ultimate microstructure of the CFO pillar. Irregular faults were observed that could be attributed to the growth of the CFO pillar on a multi-grain template island. Tilting of the CFO lattice relative to the STO substrate has also been observed and could be due to the effect of the seed island on the nucleation of the CFO pillar. These results demonstrate that the use of a template island can be highly effective in directing the growth of a matrix-pillar nanocomposite, but that the resulting pillar structure will be influenced by the microstructure of the template. Future work examining ways to make highly uniform template islands could enhance the structural uniformity of the pillars in the nanocomposite and enable the use of the materials in devices.

## Acknowledgements

Research performed in part at the National Institute of Standards and Technology (NIST) Center for Nanoscale Science and Technology. The authors would also like to thank Richard Kasica of NIST for assistance with electron-beam lithography for the patterned nanocomposite and Dr. Craig Johnson and Dr. Steven Spurgeon for helpful discussions of the results. The authors gratefully acknowledge funding from the Nanoelectronics Research Initiative, NSF (DMR-08-19762) and DARPA (HR-0011-10-1-0072). SAW and JL also thank Virginia Innovation Partnership (VIP) funding as part of the U.S. Department of Commerce's i6 Challenge. RC also wishes to acknowledge funding from the National Defense Science and Engineering Graduate Fellowship.

## Notes and references

[a] University of Virginia, Department of Materials Science and Engineering, Charlottesville, VA 22904

[b] National Institute of Standards and Technology, Center for Nanoscale Science and Technology, Gaithersburg, MD 20899, USA

[c] University of Virginia, Department of Physics, Charlottesville, VA 22904

[+] Corresponding author: rcomes@virginia.edu; Present address: Pacific Northwest National Laboratory, Fundamental and Computational Sciences Directorate, Richland, WA 99354

Electronic Supplementary Information (ESI) available: details on the film growth conditions, Cross-sectional transmission electron microscope (TEM) sample preparation methods, TEM analysis methods employed for dislocation analysis.

# Microstructural Effects of Chemical Island Templating in Patterned Matrix-Pillar Oxide Nanocomposites

**Ryan Comes[*], Kerry Siebein, Jiwei Lu and Stuart A. Wolf**

Dr. Ryan Comes
University of Virginia, Department of Materials Science and Engineering, Charlottesville, VA 22904
Present address: Pacific Northwest National Laboratory, Fundamental and Computational Sciences Division, Richland, WA 99352
E-mail: rcomes@virginia.edu

Dr. Kerry Siebein
National Institute of Standards and Technology, Center for Nanoscale Science and Technology, Gaithersburg, MD 20899, USA

Prof. Jiwei Lu
University of Virginia, Department of Materials Science and Engineering, Charlottesville, VA 22904

Prof. Stuart A. Wolf
University of Virginia, Department of Materials Science and Engineering & Department of Physics, Charlottesville, VA 22904

**Fabrication Methods and Materials Properties**

$CoFe_2O_4$-$BiFeO_3$ (CFO-BFO) nanocomposite films for this work were grown via pulsed electron deposition (PED, Neocera, Inc.)[1] using a technique that has been described previously.[2] The patterned sample was produced using a directed self-assembly process that has also been described previously.[3] A detailed description of the pulsed electron deposition chamber and process employed in our research group has also previously been published.[4] The reader is referred to these works for an understanding of the growth kinetics of the PED process. An initial film of pure $CoFe_2O_4$ (CFO) was grown on Nb-doped $SrTiO_3$ from a stoichiometric CFO target. The film showed a uniform island distribution with thickness of 12.5 nm. This sample was then patterned using the techniques outlined in Ref. 3 to produce the island template shown in Figure 1 of the main paper. The patterned substrate was then loaded into the PED chamber and a CFO-BFO nanocomposite film was grown on the substrate to produce the patterned sample. The growth conditions for these two samples are shown below in Table 1. The temperature of the substrate was determined via a calibration curve obtained using a thermocouple mounted to a sample holder and swept across the set temperatures of the resistive sample heater.

| Film | Set Temperature (°C) | Calibrated Substrate Temperature (°C) | Operating Pressure (Pascal) | Gas Composition | Pulse Voltage (kV) | Pulse Rate (Hz) | Number of Pulses (thousands) |
|---|---|---|---|---|---|---|---|
| $CoFe_2O_4$ | 700 | 515 | 1.6 | 100% $O_2$ | 8 | 8 | 40 |
| $CoFe_2O_4$-$BiFeO_3$ Composite | 775 | 577 | 2.1 | 100% $O_2$ | 11.5 (CFO), 11.8 (BFO) | 2.5 (CFO), 5 (BFO) | 37.5 (CFO), 75 (BFO) |

**Table 1: Growth conditions for sample analyzed in paper.**

**TEM Sample Preparation**

To prepare the samples for TEM analysis, a dual-beam scanning electron microscope (SEM)/focused ion beam (FIB) system was used to extract a cross-sectioned lamella. An FEI Helios 650[1] system was used for this work with a Ga ion source and a field emission electron gun. The system is equipped with several gas injection system (GIS) sources, which are used to inject a variety of different precursor metal-organic gases for deposition. To deposit these materials, either the SEM or FIB gun is used to crack the precursor gas on the surface of the sample. The system is also equipped with an Oxford Omniprobe[1] sample manipulator, which is used during the TEM sample preparation process to lift lamella from the substrate.

Figure S1 shows the progression of the lamella extraction process from a patterned nanocomposite. For the patterned nanocomposite, initial 1 μm tall, 1 μm diameter Pt metallic pillars were deposited around the array of interest for use as alignment marks using the electron gun with a beam current of 800 pA and 5 kV accelerating voltage. Both patterned and unpatterned nanocomposites were then coated with approximately 100 nm of amorphous carbon using a Gatan Precision Etching and Coating System.[1] The carbon serves as a conductive coating to reduce the effects of charging that occur due to the insulating nanocomposite film. The samples were then placed in the dual-beam system for the lift-out process. A 2 μm thick initial Pt rectangle with length of 20 μm and width of 2 μm was deposited along the <110> surface axis using the ion beam source with 30 kV and 0.23 nA beam conditions. This serves as a protective coating during the extraction process. An image of the rectangle on the patterned sample is shown in Figure S1(A). The patterned alignment marks to find the arrays are visible in the image as faint lines of contrast on the surface of the film. The FIB gun is then used to etch a trench into the substrate surrounding the Pt rectangle, with gun conditions set to 30 kV, 9.3 nA. A reduced beam current of 2.5 nA is used to clean any residual material from the trench. The sample is then tilted to undercut the substrate beneath the Pt rectangle. Using the Omniprobe manipulator, a micron scale tip is then mounted to the lamella by depositing Pt using the ion beam source with a beam current of 24 pA at the interface between the probe and lamella. This step is shown in Figure S1(B). With the probe attached, the lamella is then cut from the substrate using the FIB with a beam current of 2.5 nA. The sample is then attached to a Cu TEM grid by depositing Pt with the ion beam source with 80 pA current. The probe is then cut free using the FIB, leaving the lamella attached to the grid. This is shown in Figure S1(C). Finally, a portion of the sample is progressively thinned from its initial 2 μm thickness to less than 100 nm using the ion gun with 230 pA beam current. A view of the final thinned lamella is shown in Figure S1(D).

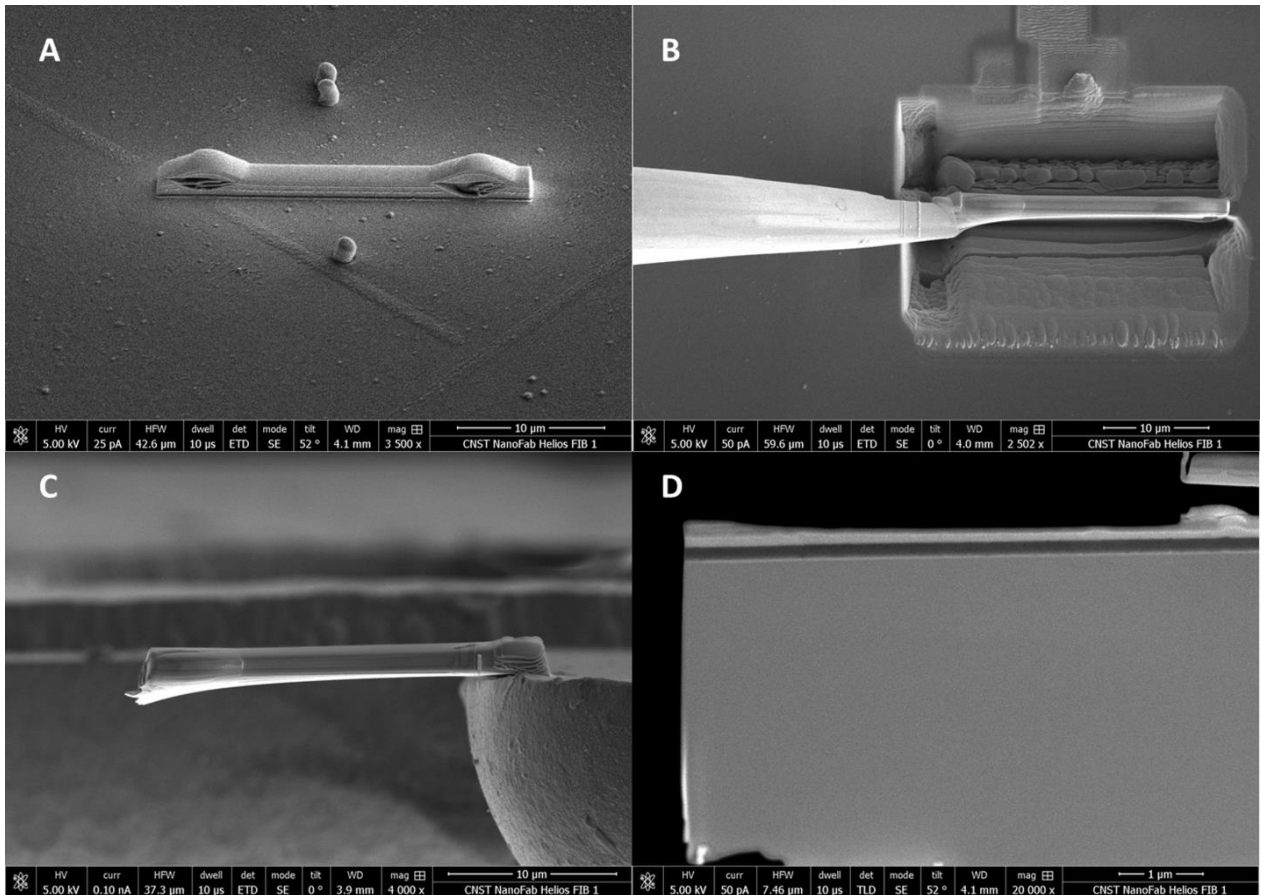

**Figure S1: Preparation process for TEM lamella. A)** Deposition of Pt protective barrier along <110> axis. **B) Trench milling and Omniprobe attachment to lamella. C) Mounting of lamella on Cu grid. D) View of final thinned lamella.**

**Faulty Templating**

During the EBL process, the effective dose in the resist is dependent on the number of nearby islands also being exposed, due to the significant number of backscattered electrons from a 100 keV electron gun. Near the edges, there are fewer nearby islands being patterned, so the effective dose to the resist is reduced. Thus, the pattern may change in some cases. In Figure S2, it is clear that the center of a separate array on the substrate received a dose that produced large diameter seed islands after etching. The STO substrate was not completely re-exposed, producing a highly defective region with no evidence of CFO pillars. At the corner of the array, ideal pillars were formed due to the reduced dose. This, along with the absence of pillars at the edges of the array in the main text suggests that chemical seeding, rather than topographic, drives the formation of CFO pillars.

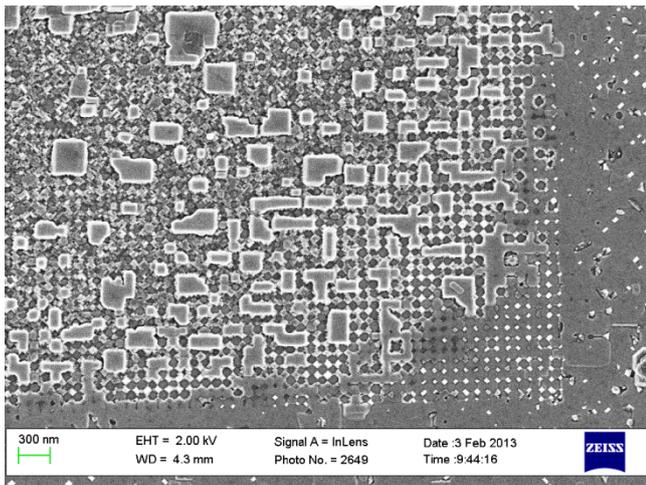

**Figure S2: SEM image of array of pillars that was overexposed in the center of the pattern but received an ideal dose at the corner.**

**Energy Dispersive X-ray Spectroscopy Analysis Techniques**

The Energy Dispersive X-ray Spectroscopy (EDS) analysis was performed with an EDAX SiLi detector. The maps were EDS drift corrected spectrum images acquired through the TEM Imaging and Analysis software interface.[1] The energy resolution is 10 eV per channel, 4 second integration time and a shaping time of 25.6 microseconds. A camera length of 0.10 meter and a beam spot size of 6 were used for eds mapping. The beam current with spot size 6 is 0.35 nA. Representative EDS spectra acquired from the substrate, matrix and pillar are shown in Figure S5. It should be noted that Bi is present along the beam path of the pillar, making the signal non-zero for the pillar in Figure S5(C). Additionally, an asterisk in Figure S5(C) denotes the neighboring Co K$\alpha$ and Fe K$\beta$ peaks , which overlap and produce non-zero Co signal in the BFO matrix. The presence of redeposited Cu from the TEM mounting grid was detected in EDS and is likely a product of the FIB sample preparation process. Likewise, we would expect trace amounts of Pt and Ga from the preparation process, but if they are present they do not rise above the noise level in the EDS spectra.

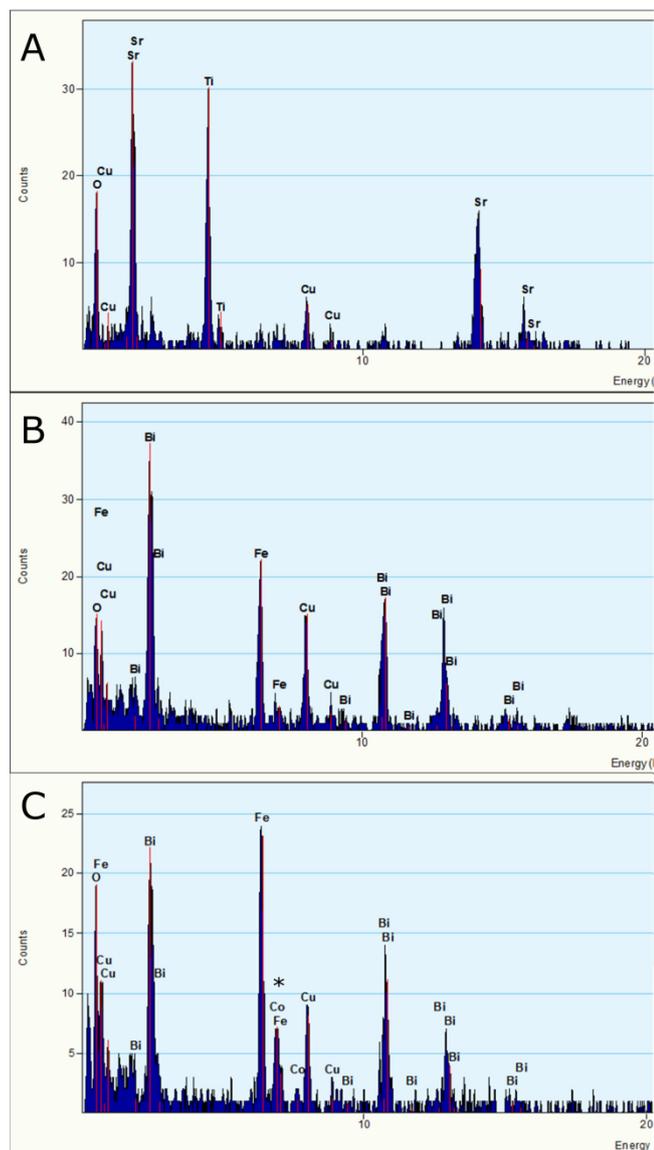

**Figure S5: EDS spectra acquired from A) Nb-doped SrTiO$_3$ substrate; B) BiFeO$_3$ matrix; C) CoFe$_2$O$_4$ pillar.**